\documentclass[useAMS,fleqn,usenatbib]{mn2e}
\usepackage{times}
\usepackage{amsmath}
\usepackage{bm}
\usepackage{url}
\usepackage{graphicx}
\usepackage{hyperref}
\usepackage{textcomp}
\usepackage{color}

\input amssym.tex

\mathchardef\mhyphen="2D

\newcommand{\bx}{\textbf{x}}

\newcommand{\avg}[1]{\left\langle{#1}\right\rangle}

\newcommand{\br}{\textbf{r}}

\newcommand{\hmpc}{\,$h^{-1}$\,Mpc}
\newcommand{\hgpc}{\,$h^{-1}$\,Gpc}

\newcommand{\hmpcnosp}{$h^{-1}$\,Mpc}

\newcommand{\LCDM}{$\Lambda$CDM}

\newcommand{\cent}{\text{\textcent}}

\definecolor{ForestGreen}{rgb}{0.3,0.7,0.3}

\newcommand{\chg}[1]{#1}

\chardef\til=`\~

\newcommand{\apj}{{ApJ}}
\newcommand{\prd}{{Phys.\ Rev.\ D}}
\newcommand{\apjl}{{ApJL}}
\newcommand{\apjs}{{ApJS}}
\newcommand{\mnras}{{MNRAS}}

\newcommand{\aap}{{A\&A}}

\newcommand{\physrep}{{Phys.\ Reports}}

\begin{document}

\title[Density-dependent BAO shifts]{Density-dependent clustering: I. Pulling back the curtains on motions of the BAO peak}

\author[M.\ C.\ Neyrinck et al.] {
{\parbox{\textwidth}{
Mark C.\ Neyrinck,$^{1,2,3}$
Istv\'{a}n Szapudi,$^4$
Nuala McCullagh,$^1$
Alexander S.\ Szalay,$^3$
\mbox{Bridget} Falck,$^5$
Jie Wang$^6$}}
\\
$^{1}$ Institute for Computational Cosmology, Department of Physics, Durham University, South Road, Durham DH1 3LE, UK\\
$^{2}$ Institut d'Astrophysique de Paris, 98 bis bd Arago, 75014; Sorbonne Universit\'{e}s, UPMC Univ Paris 6 et CNRS, UMR 7095, Paris, France\\
$^{3}$ Department of Physics and Astronomy, The Johns Hopkins University, Baltimore, MD 21218, USA \\
$^4$Institute for Astronomy, University of Hawaii, 96822 Honolulu, HI, USA\\
$^5$Institute of Theoretical Astrophysics, University of Oslo, PO Box 1029 Blindern, N-0315, Oslo, Norway\\
$^6$National Astronomy Observatories, Chinese Academy of Science, Datun Road, Beijing, PR China
}
%\author{Mark C. Neyrinck\\
%{\rm \small 
%Department of Physics and Astronomy, The Johns Hopkins University, Baltimore, MD 21218, USA}
%}

\maketitle

\begin{abstract}
The most common statistic used to analyze large-scale structure surveys is the correlation function, or power spectrum. Here, we show how `slicing' the correlation function on local density brings sensitivity to interesting non-Gaussian features in the large-scale structure, such as the expansion or contraction of baryon acoustic oscillations (BAO) according to the local density.  The sliced correlation function measures the large-scale flows that smear out the BAO, instead of just correcting them as reconstruction algorithms do. Thus, we expect the sliced correlation function to be useful in constraining the growth factor, and modified gravity theories that involve the local density. Out of the studied cases, we find that the run of the BAO peak location with density is best revealed when slicing on a $\sim 40$\hmpc\ filtered density. But slicing on a $\sim100$\hmpc\ filtered density may be most useful in distinguishing between underdense and overdense regions, whose BAO peaks are separated by a substantial $\sim 5$\hmpc\ at $z=0$. We also introduce `curtain plots' showing how local densities drive particle motions toward or away from each other over the course of an $N$-body simulation.
\end{abstract}

\begin {keywords}
  large-scale structure of Universe -- cosmology: theory
\end {keywords}

\section{Introduction}
The initial density fluctuations in the Universe are consistent with being perfectly Gaussian \citep{PlanckNonGaussianity2014}. This is an appealingly simple situation, not only physically, but statistically, since it makes the power spectrum (or its Fourier dual, the 2-point correlation function) optimal for scientific analysis of fully linear-regime observations such as the cosmic microwave background. But cosmological density fields on non-linear scales have substantial non-Gaussianity.  It is a major challenge to capture all of the scientifically-relevant information from observations of the large-scale structure of the Universe, both in principle, and in a form that enables easy analysis and understanding.

The power spectrum can still be measured from non-Gaussian fields, of course, and used to constrain cosmology, galaxy formation, or whatever the observations might depend on. But, applied to a non-Gaussian field, the power spectrum is a blunt tool, and can fail to capture the vast majority of Fisher information available in principle. Practically, this means drastically degraded parameter constraints. This is known as the information plateau: as non-linearly small scales are analyzed, error bars do not shrink as they would for a Gaussian field, due to high variance and covariance on small scales \citep{RimesHamilton2005,RimesHamilton2006,NeyrinckEtal2006,NeyrinckSzapudi2007,TakadaJain2009,TakahashiEtal2011,CarronNeyrinck2012,CarronSzapudi2014,WolkEtal2015,ReppEtal2015}.

The conventional approach to measure information beyond the power spectrum is to measure polyspectra, or higher-point correlations. But in the non-linear regime, the `full hierarchy' of these fails to contain all statistical information. \citet{Carron2011} showed that for a lognormal random field, the full information resides in the power spectrum of the log density, but at small scales, the full hierarchy of correlations can miss dramatic amounts of information. The dark-matter density field is only approximately lognormal, and initial information is lost in principle from the coarse-grained density field, e.g.\ in stream crossing. But the field seems close enough to lognormal that a logarithm renders the field nearly as suitable for analysis by the power spectrum as a Gaussian field would be, enhancing the information content in the power spectrum by orders of magnitude in an ideal measurement \citep{NeyrinckEtal2009}.

Going to lower instead of higher order, the 1-point probability density function (PDF), essentially `counts in cells', already can contain plentiful information absent from the $N$-point correlation hierarchy. A well-known example in large-scale structure of a 1-point statistic that contains information not in the first few correlation functions (indeed, often not in the full hierarchy, although this was not pointed out at the time) is the void probability function \citep{White1979}. Analyzing the full PDF instead of just its moments is essential for maximal information extraction from strong-tailed distributions like the lognormal, since the moments do not uniquely characterize it, as is well-known in statistics \citep{AitchisonBrown1957,ColesJones1991,Carron2011}. Jointly analyzing a field's 1-point PDF, together with its higher-point statistics after Gaussianization \citep{Neyrinck2014}, could even reduce some effects of biasing \citep{McCullaghEtal2016}. (Gaussianization here means applying a rank-preserving monotonic transformation to produce a Gaussian 1-point PDF.) Recent analyses of cosmological fields have included measurements of the 1-point PDF, containing information not in the power spectrum \citep{HillEtal2014,LiuEtal2016}.

Measuring the baryon acoustic oscillation (BAO) feature requires going at least to 2 point statistics. The quantity carrying the full information at 2-point order is the 2-point PDF $f(\delta_1,\delta_2;r)$. We call the (over)density $\delta=\rho/\bar{\rho}-1$, where $\bar{\rho}$ is the mean, simply `the density.' $f$ is the joint PDF of densities $\delta_1$ and $\delta_2$ (symmetric in them) at points separated by a distance $r$. The usual correlation function is an integral (Eq.\ \ref{eqn:xidef}) over $f$'s two density arguments.

An example of the high information content of the 2-point PDF is that the correlation function of any function of the density can be constructed by integrating over it with different weights. But having 3 arguments already makes the 2-point PDF possibly prohibitively unwieldy, for a few reasons. One reason is the sheer difficulty of visualizing, understanding, and modeling it. Also, in a practical measurement, slicing into the data into 3 arguments reduces the signal-to-noise ratio (S/N) of each bin of $(\delta_1,\delta_2,r)$. That is even assuming that the covariance matrix, now of 6 quantities, can be tamed. Analysis of the 2-point PDF may be more tractable than it seems, though, if the apparent near-Gaussianity of its copula \citep{ScherrerEtal2010} can be exploited. Applied to spatial statistics, the copula is a function giving all remaining information necessary to construct an $N$-point PDF from the 1-point PDF.

\subsection{Density-dependent clustering}
The statistic we introduce here, the sliced correlation function, is an integral over 1 instead of 2 of the densities in $f(\delta_1,\delta_2; r)$. Packaging the 2-point PDF into $\xi(r)$ gives something easily manageable, but at the cost of losing all sensitivity to non-Gaussian features.\footnote{Here we mean non-Gaussianity in the measured density field itself, not primordial non-Gaussianity, which can affect the ratio of the power spectrum of a biased tracer to the matter power spectrum \citep{DalalEtal2008}.} Here, we explore going halfway: integrating $f(\delta_1,\delta_2; r)$ over only one density argument. The result retains sensitivity to physically interesting non-Gaussian effects, but is manageable enough for that sensitivity to be clearly visible.

Our particular formulation of sliced correlations is new, but they are not unrelated to previous investigations of how clustering depends on galaxy properties, such as density (`environment') and luminosity. Marked correlation functions weight pairs differently based on some property of each point \citep{illian2008statistical,SzapudiEtal2000,FaltenbacherEtal2002,Sheth2005,ShethEtal2005,SkibbaEtal2006}. Density-marked correlation functions \citep{WhitePadmanabhan2009,White2016} are a bit in the spirit of sliced correlations, since changing the mark's dependence with density can probe low- and high-density regimes separately. Note that density-marked correlation functions are similar to some higher-order statistics, two-point cumulant correlators \citep[e.g.][]{SzapudiSzalay1997}, which measure two-point correlations of (usually positive integer) powers of $\delta$.

There have been previous theoretical and observational measurements of how the correlation function or power spectrum depends on the larger-scale density within a patch to regions of certain density \citep{AbbasSheth2005,AbbasSheth2007}. \citet{ChiangEtal2014,ChiangEtal2015} have looked at the density dependence of the auto-power spectrum measured in different patches (a quantity related to three-point correlations), even measuring this dependence at high significance in the Sloan Digital Sky Survey (SDSS). \citet{PujolEtal2016} have investigated some similar issues, looking at how scale-dependent halo bias depends on density. Also, \citet{UhlemannEtal2016} have developed a model, remarkably accurate into the mildly non-linear regime, of the 2-point PDF (where `points' are spheres of some separation), which could be quite useful for modeling our results analytically. This model is based on spherical-collapse dynamics and the large-deviation principle \citep[e.g.][]{BernardeauEtal2014,BernardeauEtal2015}.

Another way to investigate density dependence is through analyzing voids and superclusters separately, which has proven useful for cosmological analysis \citep[e.g.][]{GranettEtal2008,HamausEtal2016,MaoEtal2016}. One issue for studies using voids is how to define them \citep[e.g.][]{ColbergEtal2008}, or more generally, other morphologies of the cosmic web \citep[e.g.][]{LibeskindEtal2017}. It is fair to use any  procedure as long as that procedure matches in both mocks and observations, but our approach of sliced correlations may offer similar physical sensitivities as using voids and clusters, without the ambiguity attached to identifying individual structures. Note that while here we slice correlations on the density itself, one can also slice on a function(al) of the density, e.g.\ a Mexican hat wavelet that quantifies the `voidiness' or `clusteriness' on some scale at each position.

\subsection{Shifts in the baryon acoustic feature}
In this paper, we focus on how baryon acoustic oscillations (BAO) show up in sliced correlations, with features not visible in the usual correlation function. The BAO are a `standard ruler' of great importance in measuring the cosmic expansion history \citep{EisensteinEtal2005,ColeEtal2005,CuestaEtal2016,AlamEtal2017}, and detecting them drives many current and future large-scale structure surveys.

One reason the BAO are so useful is that they are essentially linear-regime, $\sim 100$\hmpc\ features. At low redshift, the peak remains nearly unmoved from its initial location in comoving coordinates, but other non-linearities are far from negligible. The largest non-linear effect is that flows on $\sim 10$\hmpc\ scales broaden the feature in $\xi(r)$ \citep{CrocceScoccimarro2006}. These flows are simple physically: in comoving coordinates, overdense regions contract, pulling the BAO peak inward, while underdense regions expand, pushing it outward \citep{SherwinZaldarriaga2012,McCullaghEtal2013}. There is also a small shift in the BAO position  \citep{SmithEtal2007,SeoEtal2010} that these motions produce, because of the large weight that overdense regions get in the usual $\xi(r)$.

Lagrangian linear theory \citep[the Zel'dovich approximation, ZA;][]{Zeldovich1970} captures the shifting and smearing of the BAO simply and rather accurately, which is why it underlies the most widely used `BAO reconstruction' methods \citep{EisensteinEtal2007,PadmanabhanEtal2012}. In these methods, the BAO are reconstructed by estimating and removing the flows. Refinements of the method, going beyond ZA, have been proposed, as well \citep[e.g.][]{MohayaeeEtal2006,FalckEtal2012logtrans,KitauraAngulo2012}. Of particular relevance for the current paper, \citet{AchitouvBlake2015} have even looked at how this BAO reconstruction behaves for different local densities of galaxies.

Such reconstruction methods are perhaps the best way to sharpen the BAO feature for detection, but in practice they are quite involved. More importantly, they discard whatever information might be in the flows themselves. As we will see, the sliced correlation function can extract this information, in a rather easily digested form.

Relevantly for the current BAO analysis, \citet{RoukemaEtal2015,RoukemaEtal2016} have detected in SDSS that the BAO feature is shifted inward by $\sim$6\% for galaxies separated by superclusters, which they present as a challenge for the standard cosmological model. Our analysis is quite different, but below, we find that there is a $\sim$5\hmpc\ difference between the BAO peak position in overdense and underdense regions assuming the standard cosmological model, broadly consistent with their measurement. Also, \citet{KitauraEtal2016} have found a BAO signal in a void catalog from SDSS Baryon Oscillation Spectroscopic Survey (BOSS) galaxies; our definition of an underdense patch is again quite different, but we find below that this is likely a good strategy, since the BAO peak position seems less biased with respect to the mean in underdense patches than overdense patches.

\section{Definitions}
The definition of the usual correlation function is
\begin{equation}
\xi(r) = \avg{\delta(\bx)\delta(\bx+\br)}=\iint \delta_1 \delta_2 f(\delta_1,\delta_2; r) d \delta_1 d\delta_2,
\label{eqn:xidef}
\end{equation}
where $f(\delta_1,\delta_2; r)$ is the 2-point PDF, the joint distribution of densities $\delta_1$ at $\delta(\bx)$, and $\delta_2$ at $\delta(\bx+\br)$. The angled brackets here denote the expectation value. We assume isotropy here, so $\xi(r)$, the covariance of the distribution, is only a function of separation.

We introduce two related quantities. The {\it contour correlation function} $\cent(r,\delta^\prime)$ is the cross-correlation of points in the density contour $\delta=\delta^\prime$ with regions a distance $r$ away from the contour:
\begin{equation}
\cent(r,\delta^\prime)=\avg{\delta(\bx+\br)}_{\delta(\bx)=\delta^\prime}=§\int \delta_2 f(\delta^\prime, \delta_2; r) d \delta_2.
\label{eqn:contourxidef}
\end{equation}
Often, it is useful to bin densities logarithmically, as we do in this paper. This would change $d\delta$ to $dA$, where $A=\ln(1+\delta)$.

The {\it sliced correlation function} is almost the same, except that inside the integral, the contour is weighted with its density. The sliced correlation function is  
\begin{align}
\xi(r,\delta^\prime)=\langle\delta(\bx)\delta(\bx+\br)\rangle_{\delta(\bx)=\delta^\prime}=\int \delta^\prime\delta_2 f(\delta^\prime,\delta_2; r) d \delta_2.
\label{eqn:slicedef}
\end{align}

For infinitesimal density bins, $\xi(r,\delta)=\delta\cent(r,\delta)$. But a realistic measurement involves finite density bins. Explicitly using distance and density bins $(r_i,\delta_j)$,
\begin{align}
\cent(r_i,\delta_j)&=\langle\delta(\bx+\br)\rangle_{|\br|\in r_i,\delta(\bx)\in\delta_j},\\ \nonumber
\xi(r_i,\delta_j)&=\langle\delta(\bx)\delta(\bx+\br)\rangle_{|\br|\in r_i,\delta(\bx)\in\delta_j}.
\end{align}

We focus on sliced correlation functions in this paper, but $\xi$ and $\cent$ both have their advantages. $\xi(r,\delta)$ has a more direct interpretation for those familiar with $\xi(r)$: it is the contribution to $\xi(r)$ from density $\delta$,
\begin{equation}
\xi(r) = \int{\xi(r,\delta) d\delta}.
\label{eqn:xisum}
\end{equation}
For $\xi(r,\delta)$, this sum can reconstruct the total $\xi(r)$ even with finite density bins; note that $\delta^\prime$ is inside the integral in Eq.\ ({\ref{eqn:slicedef}}). The contour correlation function $\cent(r,\delta)$ does not have this benefit, but it is sensitive to clustering even at the $\delta=0$ contour, whereas $\xi(r,\delta=0)=0$, by definition.

While these definitions are unambiguous in principle, there are some practical choices to make when using $\xi(r,\delta)$. The full $\xi(r)$ can be measured directly from an unsmoothed field, but $\xi(r,\delta)$ depends on how the density is estimated (i.e., smoothed). This can be important, as we show below when slicing $\xi$ with densities estimated with different smoothings. Density estimation can be with a fixed kernel (such as a Gaussian or top hat, which we use below), or with an adaptive method such as a Delaunay or Voronoi tessellation \citep{SchaapVdw2000,vdwSchaap2009}. With angular masks, and inhomogeneous sampling in the line-of-sight direction, estimating the density at each galaxy can be non-trivial. But as \citet{White2016} points out, most modern surveys large enough for precision BAO measurement already incorporate a careful local-density estimate, for BAO-reconstruction purposes. Even using $\delta$ itself is a choice; a useful alternative could be a function of $\delta$ such as the log-density.

In this paper, we estimate $\xi(r,\delta)$ using a density field defined on a fixed Eulerian, Cartesian grid, computing it with a fast Fourier transform (FFT) \citep{SzapudiEtal2005}. In a pair-counting method, sliced correlations can be measured by binning pairs into a two-dimensional set of bins, of separation and the density of both points in the pair. The random pairs must then be scaled by the number of random pairs expected form the 1-point density PDF. Using a $DD/RR$ estimator \citep{PeeblesHauser1974},
\begin{equation}
\xi(r,\delta)=DD(r,\delta)/[P(\delta)^2RR(r)] -1,
\end{equation}
where $DD(r,\delta)$ is the number of data-data pairs where at least one of the points is in density bin $\delta$ and separation bin $r$, $RR(r)$ is the number of random-random pairs with separation $r$, and $P(\delta)$ is the fraction of particles in density bin $\delta$.

The pair-counting and grid-FFT estimates of $\xi(r,\delta)$ should agree at large scales and high sampling, but there may be differences at scales smaller than interparticle spacings, since pair-counting only uses the density exactly at particles, whereas in a grid-FFT method, other locations are explicitly included. We plan to investigate any such differences in later work.

\section{Example: Gaussian field}
For a zero-mean Gaussian field $\delta(\bx)$, the correlation function, including the zero-lag $\sigma^2=\xi(r=0)$, entirely determines $\xi(r,\delta)$. $\sigma^2$ depends on a smoothing, which we set to the same smoothing that is used for density slicing. In this case, for all $r$, $f(\delta_1,\delta_2; r)$ is a bivariate Gaussian distribution in $\delta_1$ and $\delta_2$:
\begin{align}
f(\delta_1, \delta_2;r)&=\frac{1}{2\pi \sigma^2 \sqrt{1-\varrho^2}}\textrm{exp}\left[-\frac{\delta_1^2 + \delta_2^2 - 2\varrho \delta_1 \delta_2}{2\sigma^2(1-\varrho^2)}\right],
\end{align}
where $\varrho(r)=\xi(r)/\sigma^2$ is the correlation coefficient of $\delta_1$ and $\delta_2$. Putting this into Eq.\ (\ref{eqn:slicedef}) gives
\begin{align}
&\xi(r, \delta)=\frac{\delta^2 \varrho(r) }{\sqrt{2\pi\sigma^2}}\textrm{exp}\left(-\frac{\delta^2}{2\sigma^2}\right).
\end{align}
In terms of the usual correlation function,
\begin{align}
\xi(r, \delta)&=\xi(r)\frac{\delta^2 }{\sigma^3\sqrt{2\pi}}\textrm{exp}\left(-\frac{\delta^2}{2\sigma^2}\right).
\label{eqn:slicedcorforgauss}
\end{align}
For infinitesimal density bins, the contour correlation function $\cent(r,\delta)=\xi(r,\delta)/\delta$:
\begin{align}
\cent(r, \delta)&=\xi(r)\frac{\delta}{\sigma^3\sqrt{2\pi}}\textrm{exp}\left(-\frac{\delta^2}{2\sigma^2}\right).
\label{eqn:contourcorforgauss}
\end{align}
Note that in this Gaussian case, $\cent(r,\delta=0)=0$, but not in general for a non-Gaussian field.

\begin{figure}
    \begin{center}
    \includegraphics[width=\columnwidth]{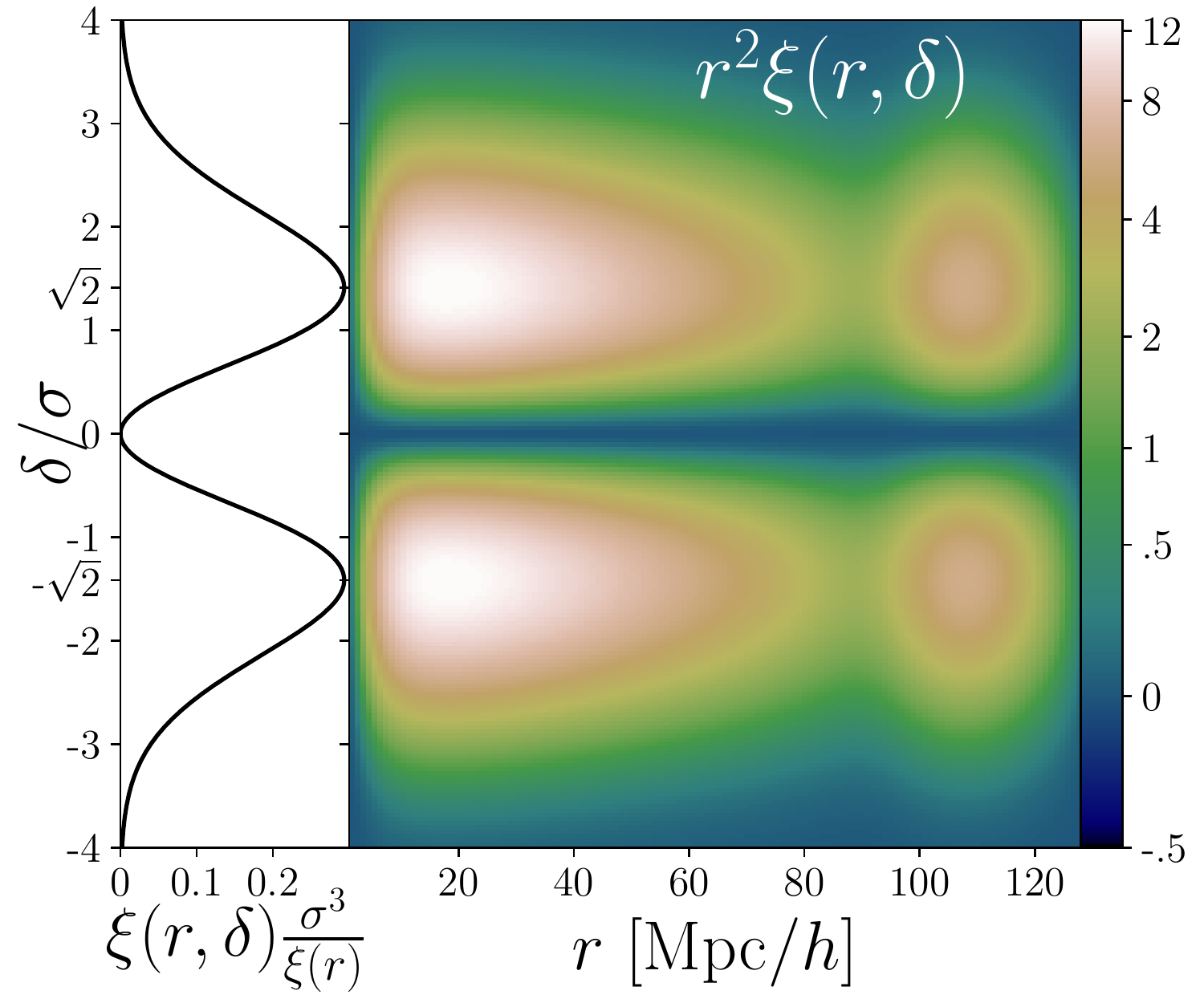}
  \end{center}  
  \caption{{\bf Right:} $r^2\xi(r,\delta)$ for a linear-theory \LCDM\ Gaussian random field, as given in Eq.\ (\ref{eqn:slicedcorforgauss}). The BAO feature comes out as blobs at the peak scale of $\sim108$\hmpc, at the base of each sideways exclamation mark. Note the nonlinear ($\sinh^{-1}$) colour scale; it goes negative for similarity to later plots, but the quantity plotted remains non-negative. {\bf Left:} a cross-section at fixed $r$. For a Gaussian field, this shape is independent of $r$.}
  \label{fig:gaussianfield_sliced_xi}
\end{figure}

Fig.\ \ref{fig:gaussianfield_sliced_xi} shows an example of $r^2\xi(r,\delta)$ for a Gaussian random field with a fiducial \LCDM\ linear-theory $\xi(r)$, smoothed with a 3 \hmpc\ 3D Gaussian filter. The left panel shows a cross-section at fixed $r$; it has a symmetric, bimodal shape, with peaks at $\delta=\pm\sigma\sqrt{2}$.

\section{Sliced correlations in simulations}
\begin{figure}
    \begin{center}
    \includegraphics[width=0.93\columnwidth]{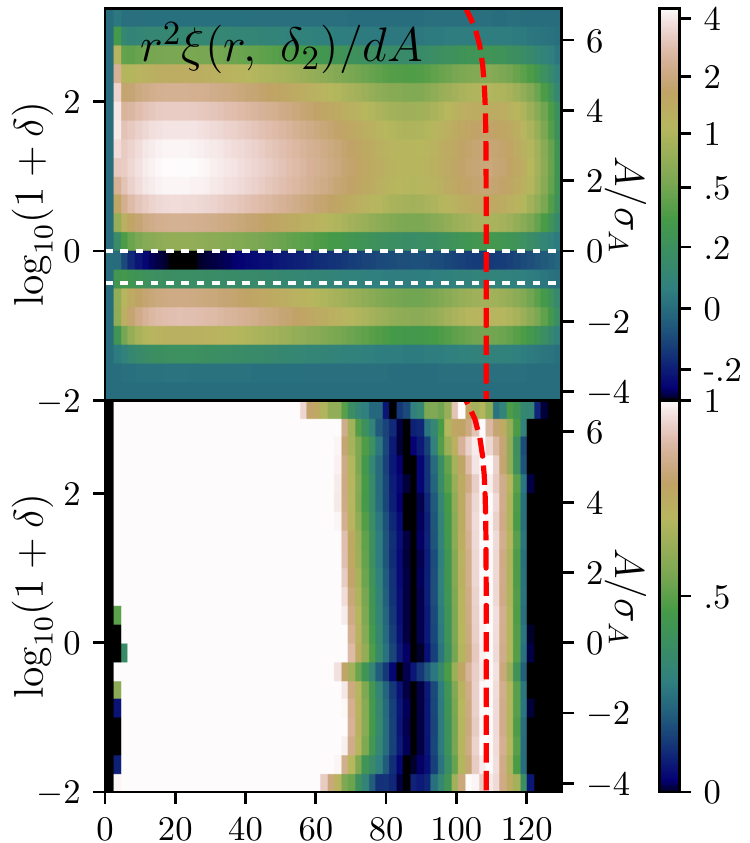}
    \includegraphics[width=0.93\columnwidth]{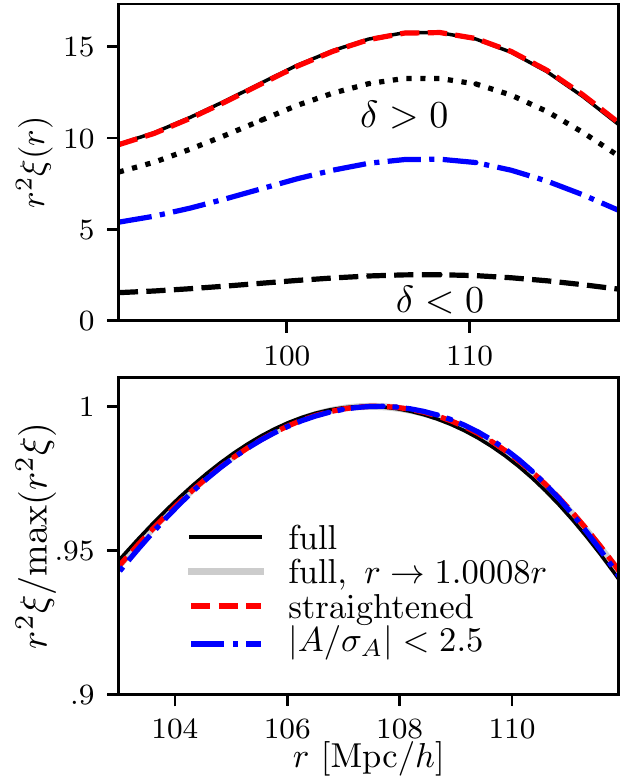}
  \end{center}  
  \caption{{\bf (1, top)}. The sliced correlation function $r^2\xi(r,\delta)$ as measured from simulations. Dotted white lines show the mean and median density. Note the non-linear colour scale. {\bf (2)} $r^2\xi(r,\delta)$, enhanced to show the BAO peak location separately in each $\delta$ bin. A dashed red curve shows a by-eye fit to the peak location in each bin, given by Eq.\ (\ref{eqn:fit}). {\bf (3)} The total $r^2\xi(r)$ (solid black), and partial sums of $r^2\xi(r,\delta)$: only $\delta>0$ (dotted); only $\delta<0$ (dashed), and bins with $A=\ln(1+\delta)$ within 2.5$\sigma$ of 0 (dotdashed blue). The red dashed curve has been `straightened,' i.e.\ $\xi(r,\delta)$ is scaled in each $\delta$ bin to straighten the dashed fit in panel 2. {\bf (4, bottom)} These curves, spline-interpolated, and normalized to unity at their maximum. The gray curve is meant to show peak motion from black to dashed red, scaling $r$ by the factor shown. In Fig.\ \ref{fig:cordens_2dcicxi_smooth40}, the various curves are more distinguishable. The average number (over simulations) of voxels in the highest two density bins is only $\avg{N_{\rm vox}}=102$ and 4, so the inward hook should be interpreted cautiously.}
  \label{fig:cordens_2dcicxi}
\end{figure}

The sliced correlations we show are averaged from $z=0$ snapshots of 218 $N$-body simulations in the Indra suite (Falck et al., in prep). Each simulation, run with {\scshape Gadget2} \citep{Springel2005}, has side length 1\hgpc, with 1024$^3$ particles, assuming a \LCDM\ cosmology: $(\Omega_m, \Omega_\Lambda, \Omega_b, h, \sigma_8, n_s)=(0.272, 0.728, 0.045, 0.704, 0.81, 0.967)$. In this paper, for simplicity as we define the statistic, we restrict our attention to the real-space matter density field at redshift zero.

We measured the sliced correlation function $\xi(r,\delta)$ in each $\delta$ bin by computing the cross-correlation of $\delta(\bx)$ with {$\left[\delta(\bx)\{\delta(\bx)\in\delta_j \}\right]$}, using an FFT. Here, $\{\delta(\bx)\in\delta_j \}$ is 1 if $\delta(\bx)$ is in the density bin $\delta_j$, and 0 if not.  This is computationally efficient if the number of density bins $\delta_j$ is small, but the computational time scales with the number of $\delta_j$. A pair-counting estimator could be more efficient than this FFT method, for a large number of density bins.

Fig.\ \ref{fig:cordens_2dcicxi} shows the real-space, dark-matter $\xi(r,\delta)$ averaged from these simulations. In the first panel, we divide each row by $dA$, the interval in $A=\ln(1+\delta)$. Thus, the total $\xi(r)$ can be reconstructed by integrating vertically. The density was estimated on a grid of 512 cells of size $\frac{1000}{512}\approx2$\hmpc, using cloud-in-cell (CIC) interpolation. The density bins are logarithmic in $(1+\delta)$, the right axis showing the log-density $A=\ln(1+\delta)$ in units of its dispersion $\sigma_A$. In the second panel, we enhance the contrast of the BAO peak in each $\delta$ bin, normalizing $\xi(r,\delta)$ to 0 at the minimum near 85\hmpc, and 1 at the peak near 108\hmpc. If $\xi(r=30$\hmpc$,\delta)<0$ at some $\delta$ (in the `valley of negativity'), we multiply $\xi(r,\delta)$ by $-1$ before enhancing the contrast.

In this and later $\xi(r,\delta)$ plots, the boundaries of the first and last density bins were determined a bit arbitrarily, to show no bins with obviously noisy features by eye in second panels, where the BAO peak is stretched to $[0,1]$. $\xi(r,\delta)$ is $\delta$ times an average density profile around a voxel with density $\delta$; thus, the noise scales with $\avg{N_{\rm vox}}$ the average number of voxels in density bin $\delta$. Interpreted this way, the fractional Poisson error (likely an underestimate) in each $(r,\delta)$ bin of $\xi(r,\delta)$ is $1/\sqrt{218 \avg{N_{\rm vox}}}$, where 218 is the number of simulations and $\avg{N_{\rm vox}}$ is the average number (over simulations) of simulation voxels in that density bin.  Except where indicated in captions, there were at minimum 1000 (out of 512$^3$) voxels on average included in each density bin, making the Poisson error quite small. In particular, features in the middle, `valley of negativity' region discussed immediately below are quite robust, drawn from over $10^7$ voxels in each simulation. Note, though, that the noise is difficult to assess in the second panels, where the procedure of stretching the peak to a range of $[0,1]$ is quite nonlinear; empirically, the noise becomes visible in second panels where $\avg{N_{\rm vox}}\lesssim100$.

Two differences from the Gaussian case are evident here, i.e.\ non-Gaussianities that $\xi(r,\delta)$ probes.

First, the full $\xi(r)$ signal is distributed among the density bins differently than in the Gaussian case. High densities greatly dominate, the contribution peaking around $\delta\sim 20$. Panel 3 shows that the contribution from $\delta<0$ (dashed) is an order of magnitude smaller than from $\delta>0$ (dotted). \chg{Second, there is even a region that contributes negative signal to $\xi(r)$, at slightly negative $\delta$}; that is, with $\xi(r,\delta)/\xi(r)<0$, a qualitative departure from the Gaussian case. We are not the first to see this sort of effect; \citet{AbbasSheth2007} found observationally that modestly underdense galaxies in SDSS are negatively correlated to the rest of the field. Also, consistent with our results, \citet{UhlemannEtal2016} found that in the low-$z$ matter density field, the $\delta$ with zero `bias' (correlation) is slightly negative; they also provided a perturbative estimate of the offset from zero.

How should we understand this `valley of negativity?' Suppose that $\delta$ is a local monotonic biasing transform $b$ of a Gaussian field $G$, $\delta = b(G)$. $\xi(\delta,r)$ generally has two classes of zero-crossings: where $\delta=0$ (by definition), but also where $\cent(\delta,r)=0$. In our measurements, $\cent(\delta,r)=0$ near the {\it median}, not mean ($\delta=0$) density. In the $\delta=b(G)$ model, $b$ transforms mean-density regions of the Gaussian $G$ to the median density of $\delta$, so in $\delta$, median-density regions are clustered like mean-density regions of a Gaussian field. Between these zeroes at the mean and median, $\xi(\delta,r)$ generally goes negative. The behaviour of $\xi(\delta,r)$ as $\delta$ varies at fixed $r$ seems driven by the non-Gaussianity of the 1-point PDF. Indeed, we expect departures from the Gaussian $\xi(r,\delta)$ to increase as the density is smoothed on smaller scales, giving a more non-Gaussian 1-point PDF. We plan to explore this behaviour in an upcoming paper.

\subsection{BAO peak motions depending on small-scale density}
The second interesting non-Gaussianity, which we focus on here, is the inward motion of the BAO at high density. This effect has a clear physical origin: a patch exceeding the mean density contracts in comoving coordinates, while an underdense patch expands in comoving coordinates.

However, this effect is quite small in Fig.\ \ref{fig:cordens_2dcicxi}, only clearly evident in the densest bin, subject to substantial noise. This is because it is the density on scales comparable to the BAO scale, $\sim 100$\hmpc, that drives the feature in or out, while here, we slice on the density estimated on only 2\hmpc\ scales. This 2\hmpc\ density is only weakly correlated with the density on 100\hmpc\ scales, but evidently is correlated enough that the highest 2\hmpc\ densities tend to be in BAO-scale overdensities, that contract. Oddly, there is also a suggestion of inward motion in the lowest-density bin.

In the red dashed curve, we show a by-eye fit to the dependence of the peak location on density, formulated rather arbitrarily as
\begin{equation}
s(\delta) = 1-s_\delta\delta-s_A A.
\label{eqn:fit}
\end{equation}
Here, $A\equiv\ln(1+\delta)$, and $s_\delta$ and $s_A$ are parameters that can change with how the density is sliced. $s_\delta$ captures the inward hook at extreme high densities. $s_A=0$ in Fig.\ \ref{fig:cordens_2dcicxi}, but it is nonzero in later figures, necessary there to capture a more gradual density dependence. In Fig.\ \ref{fig:cordens_2dcicxi}, $s_\delta=3\times10^{-5}$. Generally, varying $s_\delta$ and $s_A$ by a factor of $\sim$2 from their stated values still yields reasonable by-eye fits to second panels of Figs.\ \ref{fig:cordens_2dcicxi}-\ref{fig:cordens_2dcicxi_smooth40}.

In the bottom two panels, we zoom in on the BAO feature and show partial sums of $\xi(r,\delta)$. The dashed red curve is a `straightened' $\xi_{\rm straight}(r)=\int\xi(s(\delta)r,\delta)d\delta$. This lines up the BAO at the same scale, under the assumption that the comoving expansion or contraction at $\delta\ne0$ scales $\xi(r,\delta=0)$ with the factor $s(\delta)$. When totaling up $\xi_{\rm straight}$, we scaled $\xi(r,\delta)$ by linear interpolation in $r$. Comparing $\xi_{\rm straight}$ to the full, raw $\xi(r)$ allows an estimate of how much the flows in different density regimes shift the total peak location. To measure how much of the BAO shift can be undone in principle with this straightening, we show $\xi(r)$, with $r$ scaled by a factor of $1.0008$. This is an order of magnitude less than the full $\sim0.6$\% shift described below; thus, it appears that slicing the correlation function on a small-scale density estimate captures little of the flows we are after.

The blue dot-dashed curve excludes only $>2.5\sigma$ extremes of the log-density distribution from the integral in Eq.\ (\ref{eqn:slicedef}); this also partially restores the peak position. In effect, clipping these extremes in the tail of the distribution out of the summed $\xi(r)$ is similar to clipping the density field \citep{SimpsonEtal2011}, a technique which brings low-order statistics into better agreement with perturbative predictions, and reduces covariances, providing similar benefits as Gaussianization. (In our case, clipping out cross-correlations from the total $\xi(r)$ differs from clipping the field itself; in that case, voxels over the threshold remain fully in the sample, but with a density set to the clipping threshold.) Thus, it seems that as with a logarithmic transform \citep{McCullaghEtal2013}, the reduction of the shift in the BAO peak position is another benefit of clipping. There is ambiguity in where to place the clipping threshold, but a practical approach would be to clip to a level that brings systematic errors within the statistical errors.

\subsection{BAO peak motions depending on larger-scale densities}

\begin{figure}
    \begin{center}
    \includegraphics[width=0.95\columnwidth]{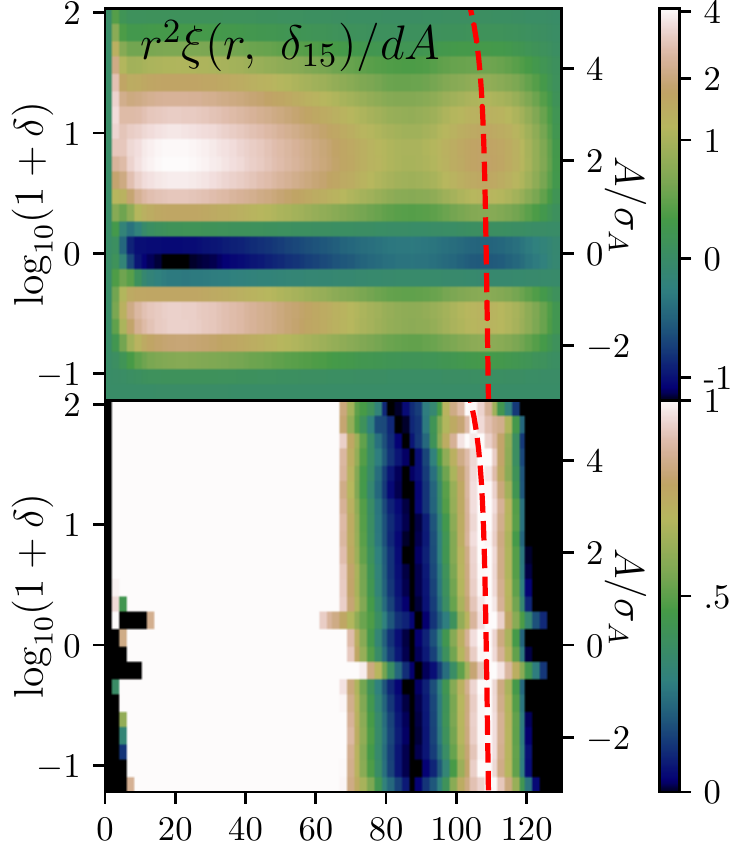}
    \includegraphics[width=0.95\columnwidth]{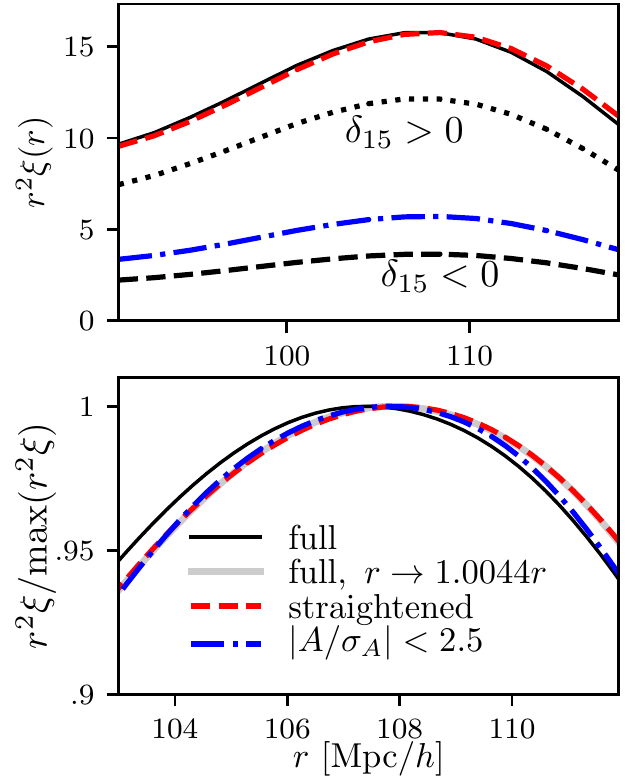}
  \end{center}  
  \caption{The same as in Fig.\ \ref{fig:cordens_2dcicxi}, except slicing on $\delta_{15}$, i.e.\ $\delta$ smoothed with a 15\hmpc\ Gaussian filter. $\avg{N_{\rm vox}}=550$ and $115$ in the highest two density bins.}
  \label{fig:cordens_2dcicxi_smooth15}
\end{figure}

\begin{figure}
    \begin{center}
    \includegraphics[width=0.95\columnwidth]{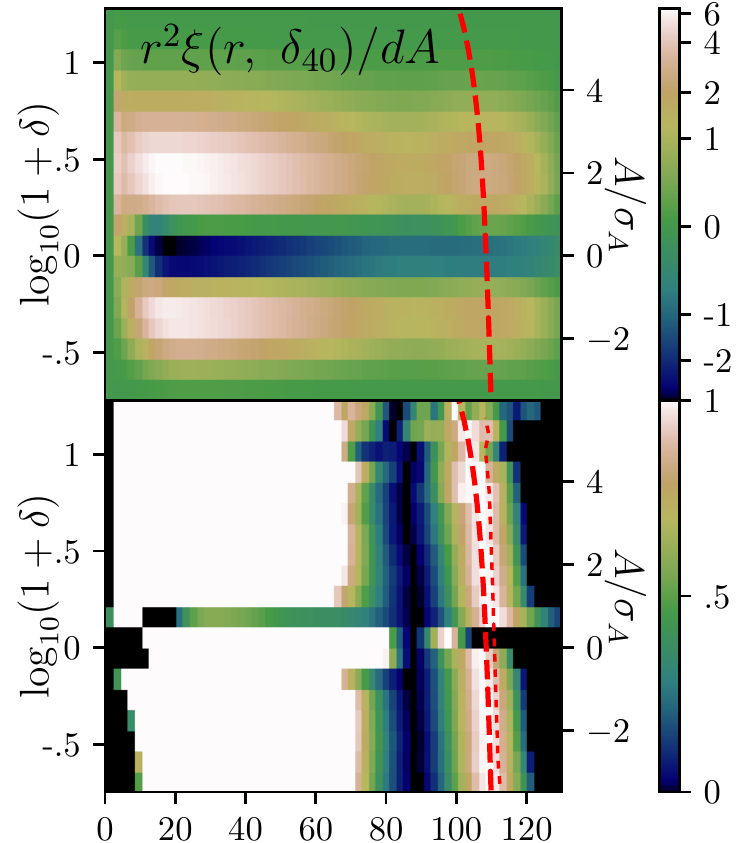}
    \includegraphics[width=0.95\columnwidth]{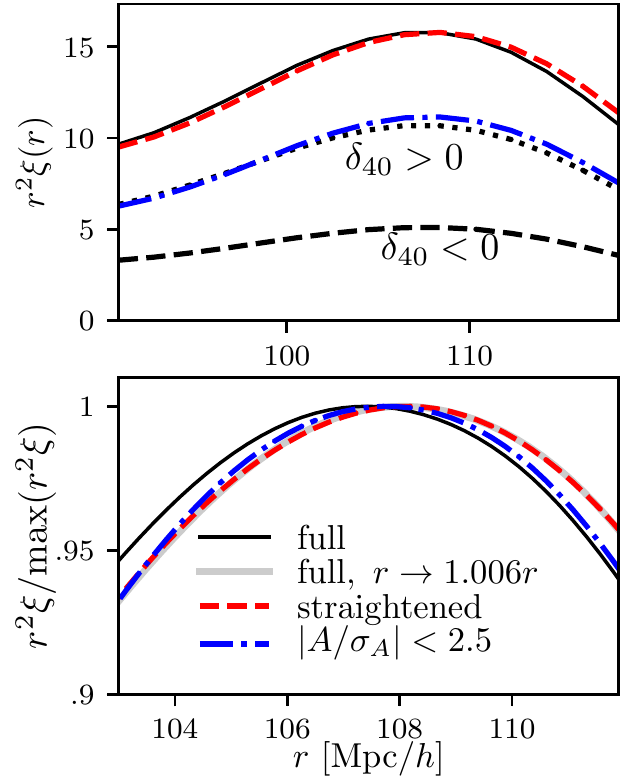}
  \end{center}  
  \caption{The same as in Fig.\ \ref{fig:cordens_2dcicxi}, except slicing on $\delta_{40}$, i.e.\ $\delta$ smoothed with a 40\hmpc\ Gaussian filter. In the second panel appears also a thinner, red dotted line, which is the blue dashed line \chg{from} the top `curtain plot' in Fig.\ \ref{fig:curtains}. This is the average radius measured \chg{from} a simulation of a shell of particles initially 105\hmpc\ away from a particle, as a function of the particle's final density, measured with this filter. The thick red line, on the other hand, was fit by eye to match the apparent behavior of the peak. The dashed and dotted red lines differ somewhat, as we discuss below in \S \ref{sec:curtains}. $\avg{N_{\rm vox}}=59$ and 289 in the highest two density bins.}
  \label{fig:cordens_2dcicxi_smooth40}
\end{figure}

While a 2\hmpc\ grid size is useful for resolving the BAO feature, we expect flows to be generated at larger scales than that. One way to capture larger-scale flows could be to measure the sliced correlations after smoothing $\delta$ on a larger scale. But we prefer to retain full resolution for positioning the peak. So, when slicing on a smoothed density field, we still measure cross-correlations from the unsmoothed (2\hmpc\ CIC) field $\delta$, but we define the {\it slices} using the smoothed field. The sliced correlation function at radius $r$ and smoothed density ${\delta_R}^\prime$ (a particular value of the density field smoothed on a scale $R$, $\delta_R(\bx)$) is
\begin{align}
\xi(r,{\delta_R}^\prime)=\langle\delta(\bx)\delta(\bx+\br)\rangle_{\delta_R(\bx)={\delta_R}^\prime}.
\label{eqn:slicesmoothdef}
\end{align}

\begin{figure}
    \begin{center}
    \includegraphics[width=\columnwidth]{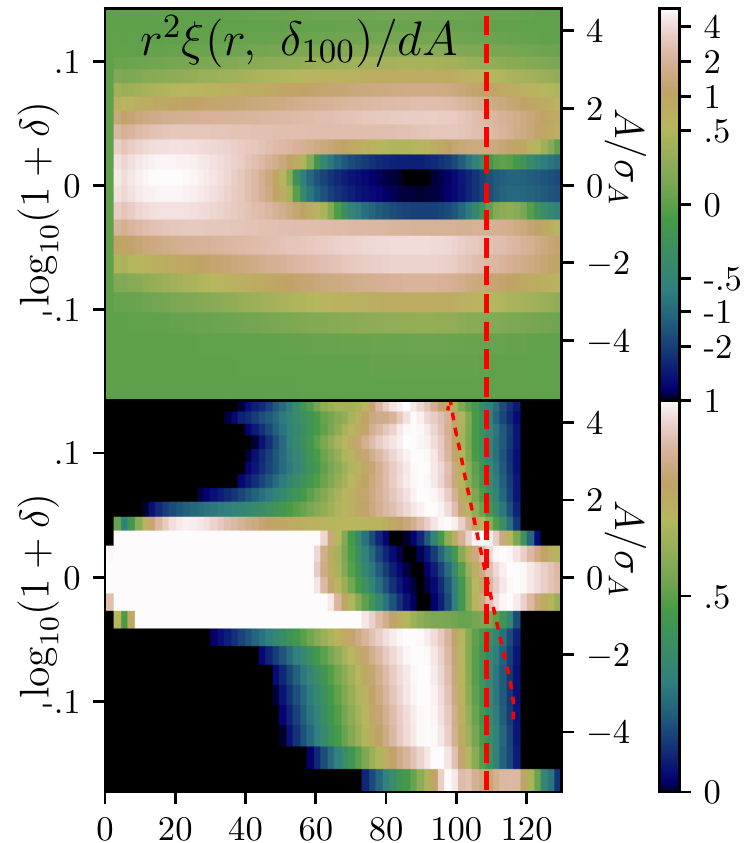}
    \includegraphics[width=\columnwidth]{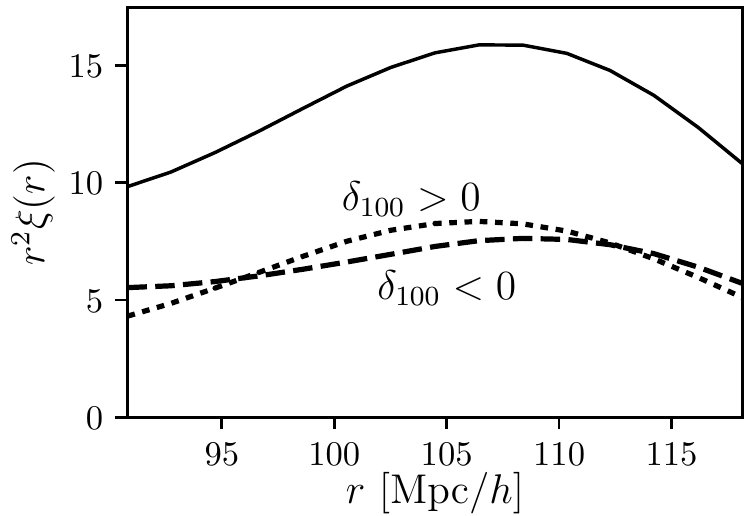}
  \end{center}  
  \caption{Similar to Figs.\ \ref{fig:cordens_2dcicxi}-\ref{fig:cordens_2dcicxi_smooth40}, except slicing on $\delta_{100}$, i.e.\ $\delta$ smoothed with a 100\hmpc\ real-space-top-hat filter. The red dashed line is a straight vertical line at the same fiducial BAO radius as in previous figures; there is little sign of BAO in the top two panels. {\bf (Middle)} At extremes in $\delta_{100}$, there are peaks at $\sim90$\hmpc, from the average profiles of regions with particular top-hat densities. This peak shows some density dependence, which could be a useful observable. The red dotted line is the blue dashed line \chg{from} the bottom `curtain plot' in Fig.\ \ref{fig:curtains}. This is the average radius measured \chg{from} a simulation of a shell of particles initially 105\hmpc\ away from a particle, as a function of the particle's final density, measured with this filter. For this filter, there is almost no correspondence between features in the sliced correlation function and the dotted red curve (which extends much farther to high $\delta$ than to low $\delta$ because it is a particle-weighted measurement). In all density bins, $\avg{N_{\rm vox}}>4000$. {\bf (Bottom)} A substantial difference is noticeable between the BAO peak location in regions that are underdense and overdense, as defined by $\delta_{100}$.}
  \label{fig:cordens_2dcicxi_smooth100}
\end{figure}

Practically, we measure the quantity in Eq.\ (\ref{eqn:slicesmoothdef}) as follows. For each density bin $j$ (a row in the top panels of Figs.\ \ref{fig:cordens_2dcicxi}-\ref{fig:cordens_2dcicxi_smooth100}), we identify the region between contours of $\delta_R(\bx)$ given by that bin's edges. We define a field that is the unsmoothed $\delta$ between these contours, and 0 outside them; we then set $\xi(r,\delta_{R,j})$ in bin $j$ to be the cross-correlation of that field with the full $\delta$ field.

A smoothing scale of $\sim$10-15\hmpc\ has been found to be optimal for Zel'dovich BAO reconstruction in galaxy samples with sparsity in the regime of BOSS \citep{PadmanabhanEtal2012,BurdenEtal2014,BurdenEtal2015,VargasMaganaEtal2017}.  While we have no expectation that smoothing on this scale is optimal for sliced correlations, it is useful to test, since it is already used in survey analysis.

In Fig.\ \ref{fig:cordens_2dcicxi_smooth15}, we slice on $\delta$ smoothed with a Gaussian of scale $15$\hmpc\ (denoted $\delta_{15}$). The dependence of the peak position with radius is now quite noticeable. The fit to the peak motion has correspondingly larger coefficients: $(s_\delta,s_A)=(2\times10^{-3},3\times10^{-4})$. The bottom panels of Fig.\ \ref{fig:cordens_2dcicxi_smooth15} show that straightening the dashed red fit to the BAO peak motion results in a much larger shift correction than in the previous section, 0.4\%-0.5\% (with an uncertainty of $\sim$0.1\%; this estimate depends on the by-eye fit of the red dashed curve).

Proceeding larger in scale, in Fig.\ \ref{fig:cordens_2dcicxi_smooth40} we slice on $\delta$ smoothed with a Gaussian of scale $40$\hmpc\ (denoted $\delta_{40}$). The fit to the peak motion has larger coefficients still: $(s_\delta,s_A)=(3\times10^{-3},6\times10^{-3})$. Straightening (removing) these motions moves the peak position outward by $\sim$0.6\%, apparently undoing the inward shift of 0.5\%-0.6\% in the real-space matter correlation function at $z=0$ found by \citet{SeoEtal2010}.

Clipping out $>2.5\sigma$ tails of the distribution from the summed $\xi(r)$ also partially restores the peak position, but not as much as `straightening.' We did not experiment with this threshold; doing so would have limited value without also considering the noise. As the threshold is brought to $\delta=0$, we would expect the peak position to be increasingly unbiased, but at the expense of increasing variance, as the volume used for analysis is decreased.

While we have enhanced the BAO motions with the larger filter, we have also exposed some curiosities. There is a bin with $\delta_{40}\approx 0$ with an apparent BAO peak at $r<100$\hmpc. We attribute this to the now-substantial difference between the unsmoothed $\delta$ field used for cross correlations, and $\delta_{40}$, the field sliced on. The $\delta_{40}\approx 0$ bin is built from a wide range of $\delta$, including contributions from the `valley of negativity,' which might interplay in a complicated way.

Also, although a density dependence for $\delta_{40}<0$ now appears, it is quite subtle. However, BAO measurements in a similar spirit, using void tracers, show more substantial differences from the full correlation function \citep{KitauraEtal2016}. Modifications to gravity at low density might be most straightforward to identify if the sliced correlation function had substantial density dependence in this regime even in GR. But sliced correlations may still be quite powerful for modified-gravity analysis: perhaps a non-GR model would carry a particular signature in this regime that our GR measurements do not have.

If the motion of the peak is enhanced with a 40\hmpc\ filter, what happens with a BAO-scale, $\sim100$\hmpc\ filter, that should entirely capture BAO-scale regions undergoing contraction or expansion with density?

Fig.\ \ref{fig:cordens_2dcicxi_smooth100} shows $\xi(r,\delta_{100})$, slicing on $\delta_{100}$, $\delta$ smoothed with a top hat of radius 100\hmpc. At density extremes, the main effect of slicing with $\delta_{100}$ seems to be the downturns at the edges of the top hat, at 100\hmpc. Perhaps this should not be surprising; if the density within a top-hat sphere is extreme, at the edge, the density will typically go toward the opposite extreme.

Oddly, the top two panels show little sign of BAO features. However, there is an interesting peak at $\sim90$\hmpc\ for extreme $\delta_{100}$. This peak is at smaller separation at high densities, and larger separation at low densities. Near $\delta_{100}=0$, there is a trough that behaves similarly. This trough may be related to the peak, carrying a negative sign as in the `valley of negativity' in previous plots.\footnote{Previously, before stretching the contrast in second panels, we flipped the sign of $\xi(r,\delta)$ in rows obviously carrying a negative sign; here, we did not do so, since the behaviour is not as clear. In each $\delta_{100}$ row, the `0' color is simply the minimum of $r^2\xi(r,\delta_{100})$ over the range 70\hmpc\ $<r<$ 120\hmpc, and `1' is the maximum over this range.} This density-dependent feature is complicated, but it may generally be produced just inside the radius of any top-hat filter used for slicing correlation functions. If so, it could offer a probe of gravitationally induced motions at each top-hat smoothing radius, unrelated to the BAO. But we leave further investigation of this possibility to future work.

The full run of $\xi(r,\delta_{100})$ with $\delta_{100}$ seems difficult to interpret, but a simple split at $\delta<0$ and $\delta>0$ gives interesting and promising results. In the bottom panel, the peaks of the $\delta_{100}>0$ and $\delta_{100}<0$ curves are a substantial $\sim 5$\hmpc\ apart.

Also, interestingly, the BAO peak has a bit higher amplitude in overdense regions; the 1-point PDF of $\delta_{100}$ is Gaussian enough ($\delta_{\rm median}=-0.003\sigma$) that this is another clear non-Gaussian feature. This may indicate some thinning and sharpening of the BAO shell itself, or an increased growth rate, in overdense regions. Applied to galaxies, the peak heights in overdense and underdense regimes may also hold information about halo bias.

We also did try a 50\hmpc\ Gaussian smoothing, giving results similar to $\delta_{40}$, but with more complicated behaviour near $\delta=0$, approaching the $\delta_{100}$ case. The optimal filter size and shape of the filter, and number of density bins (perhaps only 2), will be guided by practical considerations in an observed survey, so we stop the current study of this issue here. It is clear, though, that the correlation function is most usefully sliced using a larger smoothing radius than is typically used for BAO reconstruction.

\section{Curtain plots of density-dependent flows}
\label{sec:curtains}

\begin{figure}
    \begin{center}
     \includegraphics[width=\columnwidth]{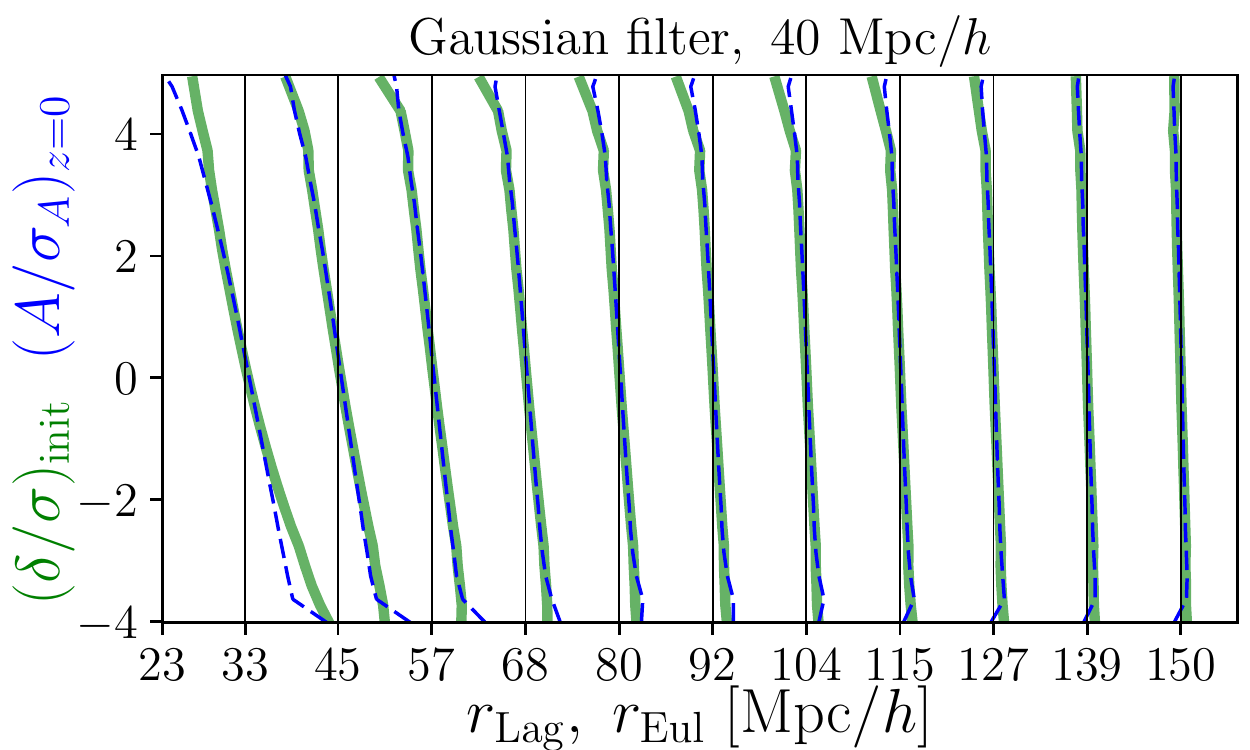}
    \includegraphics[width=\columnwidth]{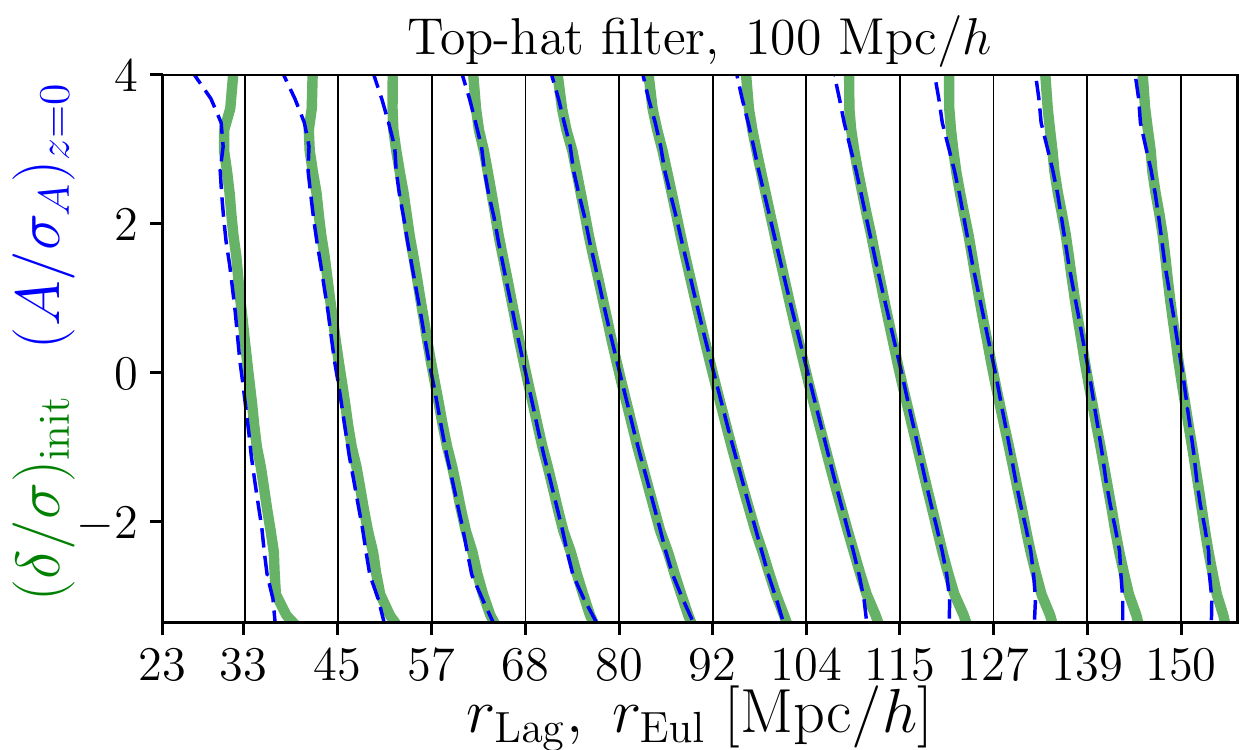}
  \end{center}  
  \caption{`Curtain' plots showing how the average inward or outward comoving motion of shells around a particle depends on that particle's density. Black lines show initial radii $r_{\rm Lag}$ of shells. Green curves show these shells' average final radii, as a function of initial central-particle density $\delta_{\rm init}$. Dashed blue curves show $z=0$ radii as a function of the log of the smoothed $z=0$ density, $A_{\rm z=0}$. BAO-radius shells around extremes of the density move by over 10\hmpc, if smoothing with a 100\hmpc\ top-hat filter.}
  \label{fig:curtains}
\end{figure}

How much do densities drive motions, in principle?  Observationally, density-dependent motions are measurable using a marker such as the BAO. But knowing all initial and final particle positions in an $N$-body simulation allows us to straightforwardly measure `curtain plots' (Fig.\ \ref{fig:curtains}). These show how the average radial motion of matter shells around a particle in an $N$-body simulation depends on the central particle's density.

For these plots, we averaged final separations $r_{\rm Eul}$ between pairs of particles in a two-dimensional histogram of initial particle separation, $r_{\rm Lag}$, and densities of one particle in the pair (each pair contributes twice; once at each particle's density). Along the $x$-axis are 4-\hmpcnosp\ bins of initial separation $r_{\rm Lag}$; along the $y$-axis are bins of initial density ($\delta_{\rm init}$) or final log-density ($A_{z=0}$), normalized by their standard deviations. Cosmic variance is small for this measurement; a single 1\hgpc, $512^3$-particle simulation sufficed, although the most extreme bins still appear to carry a bit of noise. The thin, vertical lines show shell radii in the initial conditions, before the particles move. The slanted, coloured curves, one per initial radius bin, show the final-conditions average distance to these particles (i.e., the final average radii of the shells). In underdense regions (at the bottom of the plots), particles move away from each other on average, and thus have greater Eulerian than Lagrangian separation. Likewise, in overdense regions, particles move toward each other on average, producing a smaller Eulerian than Lagrangian separation.

The scale used for filtering seems to be the scale at which motions are best picked up. In the top panel, the thick lines are most slanted at $\sim 40$\hmpc, getting more vertical at smaller and larger $r$. Similarly, in the bottom panel, the thick lines are most slanted at $\sim 100$\hmpc. So it seems unnecessary to try an even larger scale than 100\hmpc\ to detect BAO peak motions. Motions on other scales are still visible, though, because of correlations in the density and displacement field between those scales and the scale used for the filter. Note that the shapes of the filters (Gaussian or top hat) influence the plots, as well.

Binning by initial or final (log-)density makes strikingly little difference in these plots. They were estimated as follows: the initial density was reconstructed from the initial displacement field assuming the Zel'dovich approximation, and then smoothed with the filter. The final density was nearest-grid-point estimated, and then smoothed with the filter; all particles ending up in a cell were assigned the cell's smoothed density. The similarity between initial and final density may break down for smaller smoothing radius, approaching typical $\sim$10 \hmpc\ displacements. This similarity also likely relates to the usefulness of the final-conditions log-density for estimating the displacement-divergence for BAO reconstruction \citep{FalckEtal2012logtrans}. Refreshingly simply, the zero-displacement density is at $\delta=0$ in the initial and even the final conditions.

We also overplot measurements from this section in Figs.\ \ref{fig:cordens_2dcicxi_smooth40} and \ref{fig:cordens_2dcicxi_smooth100}. Specifically, we overplot the dashed blue curves from the 104\hmpc\ bin from Fig.\ \ref{fig:curtains}, scaled to 105\hmpc\ for clarity. The behaviour of the BAO peak with density does not obviously match these curtain curves, although there is qualitative agreement in the $\xi(r,\delta_{40})$ case. Evidently, other effects need to be considered for precision modeling of the BAO peak in $\xi(r,\delta_{40})$, such as the finite width of the BAO peak, and the density-dependent pileup of matter shells within it.

\section{Conclusion}
We introduce the sliced correlation function, a two-point statistic that measures density-dependent clustering. It is sensitive to several physically interesting effects that the usual correlation function misses, but we focus attention here on shifts with density of the baryon acoustic oscillation (BAO) feature.

The density-dependent motions of the BAO peak come out clearly: it moves by up to $\sim 10$\hmpc\ for the highest-density peaks, when smoothing on a 40\hmpc\ scale. These motions are sensitive to cosmological parameters:  in the Zel'dovich approximation, they are exactly scaled with the growth factor (sensitive to parameters like $\sigma_8$, $\Omega_{\rm M}$, and $w$). Since the sliced correlation function measures density-dependent clustering, it should be sensitive to modified gravity models that deviate from general relativity only in regions of low density or low gravitational potential \citep[for a review, see][]{JoyceEtal2015}. The full sliced correlation function likely has this sensitivity, but it would likely be most detectable in the density-dependent shift of the BAO peak.

We have shown examples of sliced correlation functions sensitive to BAO peak motions. Smoothing the field used for slicing (keeping the field used for cross-correlations unsmoothed) can enhance the density-dependence. For some filters (e.g.\ a 100\hmpc\ top hat), it seems that the filter itself can produce features in the sliced correlation function, obscuring BAO peaks in a finely-sliced correlation function. Still, the BAO peak positions differ by a substantial $\sim$5\hmpc\ in underdense and overdense regions as defined by this filter. But to analyze the full run of the BAO peak position with density, we suggest a $\sim$40\hmpc\ scale and a Gaussian filter shape, large enough to be sensitive to large-scale flows, but without substantial features from the filter itself. Conveniently, these filters are large enough to expect observational effects such as galaxy discreteness and bias to be manageable. Still, careful modeling will be necessary for precision analysis. We expect a 40\hmpc\ Gaussian filter to be within a factor of $\sim$2 of the optimal radius, depending on the precise question being asked, but did not exhaustively explore this question. Practically, the optimal filter size and shape may depend on survey properties such as discreteness, bias, and redshift range. Concerning filter shape, for example, spatial compactness is likely advantageous, given inevitable survey boundaries and holes.

We found some interesting features in the sliced correlation function, introduced by the shape of the filter used for smoothing and slicing. They can obscure the full run of the BAO position with density, but could be useful for other purposes. Specifically, a peak just inside the filter radius appears when slicing the correlation function using a 100\hmpc\ top-hat filter. The peak shows density dependence not present in the initial Gaussian field, and so it carries some information about how gravity has built the final density field. Possibly, such a peak arises for a top-hat filter of any radius, whose density-dependent location carries yet-untapped information. It would be interesting to find the optimal filter shape for this effect. Note that a filter could involve a field or functional estimated from the density, such as the potential.

There is much room for further study of density-dependent BAO analysis. We need to investigate how to optimize the detection of peak motions, in light of observational effects. Here, we have glossed over error bars, but as we find in preliminary work in preparation, sliced correlations of a 1-point-Gaussianized density field have increased S/N in each bin. Another question is how many slices to make. The fully sliced $\xi(r,\delta)$ could be most information-rich in principle, but it may be most convenient and powerful to analyze a single function, or small set of functions. It also may be useful to estimate the peak location and strength at each location \citep{ArnaltemurEtal2012}, using a wavelet method \citep[see also][]{XuEtal2010,TianEtal2011,LabatieEtal2012}. Looking at how the wavelet coefficients vary with density may offer a way to estimate peak motions without explicit density binning. 

The sliced correlation function has many other applications which we plan to explore. A curious feature is a `valley of negativity,' at slightly negative $\delta$, which subtracts signal from the total correlation function. This could relate to loss of information in the usual correlation function, since we find (again, in work in preparation) that the valley of negativity disappears when measuring the sliced correlation function of the 1-point-Gaussianized density (known to be more information-rich in general). The complete run of $\xi(r,\delta)$ with $\delta$ seems related to the 1-point PDF; because biasing also alters the 1-point PDF, sliced correlations may help to understand the bias of galaxies or other tracers. In redshift space, sliced correlations may be useful to understand or work around fingers of god, which corrupt low-density regions less than the high-density regions that dominate the usual correlation function.

\section*{Acknowledgments}
We thank Ravi Sheth for helping to spark this particular project with a suggestion to scale pairs according to local density (for the purpose of BAO-peak sharpening), and for helpful comments on a draft of the paper. We also thank Marius Cautun and Shaun Cole for useful discussions, and the referee for an unusually insightful and helpful report. MN and NM were supported at Durham by the UK Science and Technology Facilities Council (ST/L00075X/1) and the European Research Council (DEGAS-259586). MN was supported at IAP under the ILP LABEX (ANR-10-LABX-63) supported by French state funds managed by the ANR within the Investissements d'Avenir programme under reference ANR-11-IDEX-0004-02, and also by ERC Project No. 267117 (DARK) hosted by Universit Pierre et Marie Curie (UPMC) Paris 6, PI J. Silk.  MN and AS were supported at JHU by a grant in Data-Intensive Science from the Gordon and Betty Moore and Alfred P. Sloan Foundations. IS acknowledges support by NASA grants NNX12AF83G and NNX10AD53G, and NSF grant AST-1616974. BF acknowledges financial support from the Research Council of Norway (Programme for Space Research). JW was supported by NSFC grants 11390372 and 11373029, and the Pilot-B project.
\bibliographystyle{mnras}
\bibliography{refs}

\end{document}